\documentclass[11pt]{article}
\usepackage{mathrsfs}
\usepackage{amssymb}
\usepackage{amsmath}
\usepackage{amsbsy}
\usepackage{epsfig}
\usepackage{enumerate}
\usepackage{bm}

\newtheorem{theorem}{Theorem}[section]

\newtheorem{condition}{Condition}
\newtheorem{definition}{Definition}

\newtheorem{corollary}{Corollary}[section]

\def\proclaim#1{\par \bigskip\noindent {\bf #1}\bgroup\it\ }
\def\endproclaim{\egroup\par\bigskip}

\newbox\TempBox \newbox\TempBoxA

\def\pr{\textsf{P}} 
\def\ep{\textsf{E}} 
\def\Var{\textsf{Var}} 

\def\bk#1{\bm #1}
\def\bkg#1{\bm #1} 
\def\underwiggle 1{
\ifmmode\setbox\TempBox=\hbox{$ 1$}\else\setbox\TempBox=\hbox{1}\fi
\setbox\TempBoxA=\hbox to \wd\TempBox{\hss\char'176\hss}
\rlap{\copy\TempBox}\smash{\lower9pt\hbox{\copy\TempBoxA}}}
\begin{document}

\begin{center}
{ \large \bf  A NEW FAMILY OF COVARIATE-ADJUSTED RESPONSE ADAPTIVE
DESIGNS AND THEIR ASYMPTOTIC PROPERTIES}
\end{center}
\smallskip

\begin{center}
{\sc By Li-Xin ZHANG and Feifang HU}\footnote{Li-Xin Zhang is
Professor, Institute of Statistics and Department of Mathematics, Zhejiang University, Hongzhou,
China. Feifang Hu is Professor, Department of Statistics, University
of Virginia, Charlottesville, VA 22904-4135.  The research was
partially supported by NSF of China 10771192 (Lixin Zhang) and  NSF
Awards DMS-0349048 of USA (Feifang Hu). }

Zhejiang University and University of Virginia
\end{center}

\begin{abstract}
It is often important to incorporating covariate information in the
design of clinical trials. In literature, there are many designs of
using stratification and covariate-adaptive randomization to balance
on certain known covariate. Recently Zhang, Hu, Cheung and Chan
(2007) have proposed a family of covariate-adjusted
response-adaptive (CARA) designs and studied their asymptotic
properties. However, these CARA designs often have high
variabilities. In this paper, we propose a new family of
covariate-adjusted response-adaptive (CARA) designs. We show that
the new designs have smaller variabilities and therefore more
efficient.

\end{abstract}

\baselineskip 14.9pt

\section{Introduction}
\setcounter{section}{1} \setcounter{equation}{0}
\setcounter{theorem}{0} \setcounter{remark}{0}

Response-adaptive designs for clinical trials incorporate
sequentially accruing response data into future allocation
probabilities. A major objective of response-adaptive designs in
clinical trials is to minimize the number of patients that is
assigned to the inferior treatment to a degree that still generates
useful statistical inferences.
The preliminary idea of response adaptive randomization can be
traced back to Thompson (1933) and Robbins (1952). A lot of
response-adaptive designs have been proposed in literature ({\em
e.g.}, Rosenberger and Lachin 2002, Hu and Rosenberger, 2006). Much
recent work has focused on proposing better randomized adaptive
designs. The three main components for evaluating a
response-adaptive design are allocation proportion, efficiency
(power), and variability. The issue of efficiency or power was
discussed by Hu and Rosenberger (2003), who showed that the
efficiency is a decreasing function of the variability induced by
the randomization procedure for any given allocation proportion. Hu,
Rosenberger and Zhang (2006) showed that there is an asymptotic
lower bound on the variability of response-adaptive designs. A
response-adaptive design that attains this lower bound will be said
to be first order efficient. More recently, Hu, Zhang and He (2008)
proposed a new family of efficient randomized adaptive designs that
can adapt to any desired allocation proportion. But all these
studies are limit to the designs that do not incorporate covariates.

In many clinical trials (Pocock and Simon, 1975, Taves, 1974),
covariate information is available and has a strong influence on the
responses of patients. For instance, the efficacy of a hypertensive
drug is related to a patient's initial blood pressure and
cholesterol level, whereas the effectiveness of a cancer treatment
may depend on whether the patient is a smoker or a non-smoker.
Covariate-adaptive designs have been proposed to balance covariates
among treatment groups (see Pocock and Simon, 1975, Taves, 1974 and
Zelen, 1974). Hu and Rosenberger (2006) defined a covariate-adjusted
response-adaptive (CARA) design as a design that incorporate
sequentially  history information of accruing response data and
covariate as well as the observed covariate information of the
incoming patient into future allocation probabilities.

In a CARA design, the assignment of a treatment depends on the
history information and the covariate of the incoming patient. This
generates a certain level of technical complexity for studying the
properties of the design. Zhang, et al (2007) got a limit success on
CARA designs by proposing a class of CARA designs that allow a wide
spectrum of applications to very general statistical models and
obtaining the asymptotic properties to provide a statistical basis
for inferences after using this kind of designs. However, the CARA
designs in Zhang, et al (2007) often have high variabilities and
therefore are not efficient (Hu and Rosenberger, 2003). The major
purpose of this paper is to study the variability and efficiency of
CARA designs and to propose a new family of CARA designs with small
variabilities.

The paper is organized as follows. In Section 2, the Fisher
information and the best asymptotic variability are derived for a
CARA design with any given target allocation proportion. We will
find that the Fisher information and the variability depend on the
distribution of each individual response, the target function and
the distribution of the covariate. In Section 3, we propose a new
CARA design that can adapt to target any  allocation function and in
which  a parameter can be tuned such that the asymptotic variability
approaches to the best one. The design proposed by Zhang, et al
(2007) is a special case of this new design and has the largest
variability in all this kind of designs. The new design is also an
extension of the doubly adaptive biased coin design (BDCD) proposed
by Eisele and Woodroofe (1995) and Hu and Zhang (2004a). The
technical proofs are put on the Appendix.

\section{Variability and efficiency of CARA designs}
\setcounter{section}{2} \setcounter{equation}{0}
\setcounter{theorem}{0} \setcounter{remark}{0}

\subsection{ General framework of CARA designs.}
Given a clinical trial with $K$ treatments. Supposing that a patient
with a covariate vector $\bm \xi$ is assigned to treatment $k$,
$k=1,\ldots, K$, and the observed response is $Y_k$, assume that the
response $Y_k$ has a conditional distribution $f_k(y_k|\bm
\theta_k,\bm\xi)$ for given the covariate $\bm\xi$. Here $\bm
\theta_k$, $k=1,\ldots, K$, are unknown parameters, and $\bm
\Theta_k\subset \mathbb R^d$ is the parameter space of $\bm
\theta_k$.

In an adaptive design, we let $\bm  X_1, \bm  X_2,...$ be the sequence
of random treatment assignments. For the $m$-th subject,
 $\bm
X_m=(X_{m,1},\ldots,X_{m,K})$ represents the assignment of treatment
such that if the $m$-th subject is allocated to treatment $k$, then
all elements in $\bm  X_m$ are $0$ except for the $k$-th component,
$X_{m,k}$, which is $1$.  Suppose that $\{ Y_{m,k}, ~k=1,\ldots, K,~
m=1,2\ldots\}$ denote the responses such that $Y_{m,k}$ is the
response of the $m$-th subject to treatment $k$, $k=1,\ldots, K$. In
practical situations, only $Y_{m,k}$ with $X_{m,k}=1$ is observed.
Denote $\bm  Y_m=(Y_{m,1},\ldots, Y_{m,K})$. Also, assume that
covariate information is available in the clinical study. Let $\bm
\xi_m$ be the covariate of the $m$-th subject. We assume that
$\{(Y_{m,1},\ldots, Y_{m,K},\bm \xi_m), ~m=1,2,\ldots\}$ is a
sequence of i.i.d. random vectors, the distributions of which are
the same as that of $(Y_1,\ldots, Y_K,\bm \xi)$. Further, let
$\mathscr{X}_m=\sigma(\bm  X_1,\ldots,\bm  X_m)$,
$\mathscr{Y}_m=\sigma(\bm  Y_1,\ldots,\bm  Y_m)$ and
$\mathscr{Z}_m=\sigma(\bm \xi_1,\ldots,\bm \xi_m)$ be the sigma
fields  corresponding to the responses, assignments and covariates
respectively, and let $\mathscr{F}_m=\sigma(\mathscr{X}_m,
\mathscr{Y}_m, \mathscr{Z}_m)$ be the sigma field of the history. A
general covariate-adjusted response-adaptive (CARA) design is
defined by
\begin{align*}
\psi_{m+1,k}=&\pr( X_{m+1,k}=1|\mathscr{F}_m, \bm\xi_{m+1}) \\
=&\pr(X_{m+1,k}=1| \mathscr{X}_m, \mathscr{Y}_m,
\mathscr{Z}_{m+1}),\; k=1,...,K, \end{align*}
 the conditional
probabilities of assigning treatments $1,...,K$ to the $m$th
patient, conditioning on the entire history including the
information of all previous $m$ assignments, responses, and
covariate vectors, plus the information of the current patient's
covariate vector.

\subsection{CARA designs with a target.}
Let $N_{m,k}$ be the number of subjects assigned to treatment $k$ in
the first $m$ assignments and write $\bm  N_m=(N_{m,1},\ldots,
N_{m,K})$. Then $\bm  N_m=\sum_{i=1}^m \bm  X_i$. Further, let
 $N_{n,k|\bm x}=\sum_{m=1}^n X_{m,k} I\{\bm \xi_m=\bm x\}$
be the number  of subjects with covariate $\bm x$ that is randomized
to treatment $k$, $k=1,\ldots, K$, in the $n$ trials, and $N_n(\bm
x)=\sum_{m=1}^nI\{\bm \xi_m=\bm x\}$ be the total number of subjects
with covariate $\bm x$. Write $\bm\theta=(\bm\theta_1,\ldots,\bm
\theta_K)$. Because the value of $\bm\theta$ and the covariate
determinate the distributions of the outcomes, and accordingly, the
effects of each treatments,   in many cases one would like to define
a CARA design such that the "conditional" allocation proportion for
a given  covariate $\bm x$ converges to a pre-specified proportion
which is a function of $\bm\theta$ and $\bm x$. That is,
\begin{equation}\label{eq2.1}
\frac{N_{n,k|x}}{N_n(x)}\to \pi_k(\bm\theta,\bm x), \; k=1,\ldots,K,
\end{equation}
where $\pi_1(\bm\theta,\bm x)$, $\ldots$, $\pi_K(\bm\theta,\bm x)$
are $K$ known functions. We call them target allocation functions.
Examples for the choice of target functions are discussed in Zhang,
et al (2007), Rosenberger, et al (2001),  Rosenberger, Vidyashankar
and Agarwal  (2001)  and Hu and Rosenberger (2006). Recently,
Tymofyeyev, Rosenberger and Hu  (2007) developed a general framework
to obtain optimal allocation proportion for $K$-treatment clinical
trials. However, when $\pr(\bm\xi=\bm x)=0$, for example, in the
continuous covariate case,  the "conditional" allocation proportion
$N_{n,k|x}/N_n(x)$ is not well-defined because both the numerator
and denominator are zeros almost surely. As compared with
(\ref{eq2.1}), it is more meaningful to allocate each individual
patient to treatment $k$ with a probability close to
$\pi_k(\bm\theta,\bm x)$ for a given covariate $\bm x$. So we
consider a class of CARA designs with a property that
\begin{equation}\label{property}
\pr( X_{m+1,k}=1|\mathscr{F}_m,
\bm\xi_{m+1}=\bm x)\to \pi_k(\bm \theta,\bm x)\;\; a.s.
\end{equation}
The next theorem tells us that (\ref{property}) implies (\ref{eq2.1}). Write
$\rho_k(\bm\theta)=\ep\pi_k(\bm \theta,\bm \xi)$, $k=1,\ldots,K$,
 $\bm\rho(\bm\theta)
 =(\rho_1(\bm\theta),\ldots,\rho_K(\bm\theta))$
 and $\bm \pi(\bm\theta,\bm x)=(\pi_1(\bm\theta,\bm x),\ldots,\pi_K(\bm\theta,\bm x)).$

\begin{theorem}\label{theorem1} If (\ref{property}) is satisfied, then
\begin{equation}\label{eqtheorem1.2} \frac{N_{n,k|\bm x}}{N_n(\bm x)}\to \pi_k(\bm\theta,\bm x)\; a.s.
\; \text{ on the event } \; \{ N_n(x)\to \infty \}
\end{equation}
and
\begin{equation}\label{eqtheorem1.1}
 \frac{N_{n,k}}{n}\to \rho_k(\bm\theta)\; a.s.
 \end{equation}
 Here,  "$A\; a.s.$ on $B$" means that
 $\pr(B\setminus A)=0$ for two events $A$ and $B$. Further, if the density of the
 covariate is positive at $\bm x$, then
 \begin{equation}\label{eqtheorem1.3}
 \lim_{r\searrow 0} \lim_{n\to\infty}
  \frac{N_{n,k|B(\bm x,r)}}{N_n(B(\bm x,r))}=\pi_k(\bm\theta,\bm x)\;\; a.s.,
 \end{equation}
 where $N_{n,k|B(\bm x,r)}=\sum_{m=1}^n
 X_{m,k}I\{\bm\xi_m\in B(\bm x,r)\}$, $N_n(B(\bm x,r))=\sum_{m=1}^n
 I\{\bm\xi_m\in B(\bm x,r)\}$, $B(\bm x,r)$ is a ball with the center $\bm x$ and the radius $r$.
\end{theorem}

Notice, when $\pr(\bm\xi=\bm x)=0$, though the allocation proportion
$N_{n,k|\bm x}/N_n(\bm x)$ is not well-defined, (\ref{eqtheorem1.2})
is trivial because $\pr(N_n(\bm x)\to \infty)=0$.
Accurately, (\ref{eqtheorem1.2}) makes sense only in the discrete covariate case and (\ref{eqtheorem1.3})
is a version of (\ref{eqtheorem1.2}) for continuous covarites.

\subsection{Variability and efficiency.}
For response-adaptive designs which do not incorporate covariates,
Hu, Rosenberger  and Zhang (2006) found  the lower bound of the
asymptotic variability of a design, i.e., of the allocation
proportions of the design. A design is called asymptotically
efficient if its asymptotic variability attains the lower bound.
Next, we study the variability and efficiency of the CARA designs.
Suppose, given $\bm \xi$, that the response $Y_k$ of a trial of
treatment $k$ has a distribution in the exponential family, and
takes the form
\begin{eqnarray}\label{eqglm}
f_k(y_k|\bm \xi,
\bm \theta_k)=\exp\big\{(y_k\mu_k-a_k(\mu_k))/\phi_k+b_k(y_k,\phi_k)\}
\end{eqnarray}
 with link function $\mu_k=h_k(\bm \xi\bm \theta_k^{T})$,
 where $\bm \theta_k=(\theta_{k1},\ldots,\theta_{kd})$,
$k=1,\ldots, K$, are coefficients. Assume that the scale parameter
$\phi_k$ is fixed. It is easily checked that
$\ep[Y_k|\bm\xi]=a_k^{\prime}(\mu_k)$,
$\Var(Y_k|\bm\xi)=a^{\prime\prime}_k(\mu_k)\phi_k$,
$$ \frac{\partial\log f_k(y_k|\bm \xi,\bm \theta_k)}{\partial
\bm \theta_k}=\frac{1}{\phi_k}\{y_k-a_k^{\prime}(\mu_k)\}
h_k^{\prime}(\bm \xi\bm \theta_k^T)\bm \xi,$$
$$ \frac{\partial^2\log f_k(y_k|\bm \xi,\bm \theta_k)}{\partial
\bm \theta_k^2}
=\frac{1}{\phi_k}\Big\{-a_k^{\prime\prime}(\mu_k)[h_k^{\prime}(\bm \xi\bm \theta_k^T)]^2
+[y_k-a_k^{\prime}(\mu_k)]h_k^{\prime\prime}
(\bm \xi\bm \theta_k^T)\Big\}\bm \xi^{T}\bm \xi$$
and, given $\bm \xi$, the conditional Fisher information matrix is
$$ \bm  I_k(\bm \theta_k|\bm \xi)=-\ep\Big[\frac{\partial^2\log
f_k(Y_k|\bm \xi,\bm \theta_k)}{\partial
\bm \theta_k^2}\Big|\bm \xi\Big]
=\frac{1}{\phi_k}a_k^{\prime\prime}(\mu_k)[h_k^{\prime}
(\bm \xi\bm \theta_k^T)]^2\bm \xi^{T}\bm \xi.$$
 For the observations up to stage
$n$, the likelihood function is
\begin{align}\label{eqlikelihood}
L(\bm \theta)=&\prod_{j=1}^n\prod_{k=1}^K[f_k(Y_{j,k}|\bm \xi_j,
\bm \theta_k)]^{X_{j,k}}\nonumber\\
=&\prod_{k=1}^K
\prod_{j=1}^n[f_k(Y_{j,k}|\bm \xi_j,\bm \theta_k)]^{X_{j,k}}:=
\prod_{k=1}^KL_k(\bm \theta_k)
\end{align}
 with $ \log
L_k(\bm \theta_k)\propto \sum_{j=1}^n
X_{j,k}\{Y_{j,k}-a_k(\mu_{j,k})\}$,
$\mu_{j,k}=h_k(\bm \theta_k^{T}\bm \xi_j)$, $k=1,2,\ldots,K. $
 Write
 \begin{equation}\label{eqinformation1}
 \bm I_k=\ep[\pi_k(\bm\theta,\bm\xi)\bm  I_k(\bm \theta_k|\bm \xi)], \;\;k=1,\ldots, K.
 \end{equation}
Then
$$-\ep_{\bm\theta}
\left[\frac{\partial^2 \log L(\bm\theta)}{\partial\bm\theta_k^2}\right]=\sum_{j=1}^n\ep_{\bm\theta}
\left[X_{j,k}\bm  I_k(\bm \theta_k|\bm \xi_j)\right]=n \bm I_k+o(n)$$
It follows that the entire Firsher information matrix is
$$ \bm I_n(\bm \theta)=-\ep_{\bm\theta}
\left[\frac{\partial^2 \log L(\bm\theta)}{\partial\bm\theta^2}\right]=n diag( \bm I_1,\ldots,\bm I_K)+o(n). $$
Thus we obtain the following theorem.

\begin{theorem}\label{theorem2} Suppose the responses follow the generalized linear model (\ref{eqglm}) and the design satisfies (\ref{property}).
Let $\bm I(\bm\theta)=diag( \bm I_1,\ldots,\bm I_K)$. Then the
Firsher information matrix satisfies
$$\bm I_n(\bm \theta)=  n \bm I(\bm\theta)+o(n), $$
and the asymptotic variance-covariance matrix of an asymptotic efficient estimator of $\bm \theta$ is $\bm I^{-1}(\bm\theta)/n$.
\end{theorem}

The limit proportion
$\bm\rho(\bm\theta)=(\rho_1(\bm\theta),\ldots,\rho_K(\bm\theta))$
depends on both the parameter $\bm\theta$ and the distribution of
$\bm\xi$.  When the distribution of $\bm\xi$ is known, according to
Theorem \ref{theorem2}, the asymptotic variance-covariance matrix of
an asymptotic efficient estimator of $\bm\rho(\bm \theta)$ is
$\frac{1}{n}\frac{\partial \bm\rho(\bm\theta)}{\partial \bm \theta}
\bm I^{-1}(\bm\theta) \left( \frac{\partial
\bm\rho(\bm\theta)}{\partial \bm \theta} \right)^T. $ While, if the
parameter $\bm\theta$ is known, then the non-parameter maximal
likelihood estimator (MLE) of
$\bm\rho(\bm\theta)=\ep[\bm\pi(\bm\theta,\bm\xi)]$ is
$\frac{1}{n}\sum_{m=1}^n\bm\pi(\bm\theta,\bm\xi_m)$ and its
variance-covariance matrix is $  \Var\{\pi(\bm\theta,\bm\xi)\}/n.$
So, in the general case that  the parameter  $\bm\theta$ and the
distribution of $\bm\xi$ are both unknown, the asymptotic
variance-covariance matrix of an asymptotic efficient estimator of
$\bm\rho(\bm \theta)$ is $\bm B(\bm\theta)/n$, where
$$ \bm B(\bm\theta)=
\frac{\partial \bm\rho(\bm\theta)}{\partial \bm \theta} \bm
I^{-1}(\bm\theta) \left( \frac{\partial \bm\rho(\bm\theta)}{\partial
\bm \theta} \right)^T+\Var\{\pi(\bm\theta,\bm\xi)\}. $$ The
allocation proportion $\bm N_n/n$ in a adaptive design with property
(\ref{property}) will converge to $\bm\rho(\bm\theta)$ according to
Theorem \ref{theorem1}. So we can now define an asymptotically
efficient CARA design as follows.

\begin{definition}\label{definition1}
A covaraite-adjusted response-adaptive design with target function
$\bm\pi(\bm\theta,\bm x)$ is called asymptotically efficient if it
satisfies (\ref{property}) and
\begin{equation}\label{property2} n^{1/2}\big(\bm
N_n/n-\bm\rho(\bm\theta)\big) \overset{\mathscr{D}}\to N\big(\bm
0,\bm B(\bm\theta)\big),
\end{equation}
and  $\bm B(\bm\theta)$ is called the best asymptotic variability.
\end{definition}

\bigskip Zhang, Hu, Cheung and Chan (2007) proposed a
CARA design  (we refer it as ZHCC's design) by defining
$$ \pr(X_{m+1,k}=1|\mathscr{F}_m,\bm\xi_m)=\pi_k(\widehat{\bm\theta}_m,\bm\xi_{m+1}), $$
where $\widehat{\bm\theta}_m$ is the MLE of $\bm \theta$ based on
the observations up to stage $m$. It has been shown that ZHCC's
design satisfy (\ref{property}) and
$$n^{1/2}\big(\bm
N_n/n-\bm\rho(\bm\theta)\big) \overset{\mathscr{D}}\to N\big(\bm
0,\bm \Sigma(\bm\theta)\big),
 $$
 where
 $$\bm\Sigma(\bm\theta)=
2\frac{\partial \bm\rho(\bm\theta)}{\partial \bm \theta} \bm
I^{-1}(\bm\theta) \left( \frac{\partial \bm\rho(\bm\theta)}{\partial
\bm \theta} \right)^T
+diag(\bm\rho(\bm\theta))-\big(\bm\rho(\bm\theta)\big)^T\bm\rho(\bm\theta).
$$
It is easily seen that
\begin{align*}
&diag(\bm\rho(\bm\theta))-\big(\bm\rho(\bm\theta)\big)^T\bm\rho(\bm\theta)\\
=&\Var\{\pi(\bm\theta,\bm\xi)\}+\ep\left[diag(\pi(\bm\theta,\bm\xi))
-\left(\pi(\bm\theta,\bm\xi)\right)^T\pi(\bm\theta,\bm\xi)\right]
\ge  \Var\{\pi(\bm\theta,\bm\xi)\},
\end{align*}
where $\bm A\ge \bm B$ means that $\bm A -\bm B$ is non-negative
definite. Hence, ZHCC's design is not asymptotically efficient.

It is of significance to find an asymptotic efficient CARA design
for any given  target function $\bm\pi(\bm\theta,\bm x)$. In the
next section, we will propose a new class of CARA designs with an
asymptotic variability being able to approach the best one.

\section{Covariate-adjusted DBCD}
Our new design is based on the idea of the doubly adaptive biased
coin design (BDCD)  proposed by Eisele and Woodroofe (1995), and
extended by Hu and Zhang (2004a). In the scenario without
covariates, the Hu and Zhang's extension can target any desired
allocation and can approach the lower bound of the asymptotic
variability by tuning a parameter. In this section, we modify the
DBCD to incorporate covariates. For simplification, we only consider
the two-treatment case ($K=2$).

\noindent{\bf Covariate-adjusted DBCD (CADBCD):} To start, we let
$\bm \theta_0$ be an initial estimate of $\bm \theta$, and assign
$m_0$ subjects to each treatment by using a restricted
randomization. Assume that $m$ ($m\ge 2 m_0$) subjects have been
assigned to treatments. Their responses $\{\bk Y_j, ~j=1,\ldots,m\}$
and the corresponding covariates $\{\bkg\xi_j, ~j=1,\ldots, m\}$ are
observed. We let
$\widehat{\bkg\theta}_m=(\widehat{\bkg\theta}_{m,1},\widehat{\bkg\theta}_{m,2})$
be an estimate of $\bkg\theta=(\bkg\theta_1,\bkg\theta_2)$. Here,
for each $k=1,2$, $\widehat{\bkg\theta}_{m,k}
=\widehat{\bkg\theta}_{m,k}(Y_{j,k},\bkg\xi_j: X_{j,k}=1,
j=1,\ldots,m)$ is the estimator of $\bkg\theta_k$ that is based on
the observed $N_{m,k}$-size sample $\{(Y_{j,k},\bkg\xi_j):$ for
which $X_{j,k}=1, j=1\ldots,m\}$. Write
$\widehat{\rho}_m=\frac{1}{m}\sum_{i=1}^m\pi_1(\widehat{\bm\theta}_m,\bm\xi_i)$
and $\widehat{\pi}_m=\pi_1(\widehat{\bm\theta}_m,\bm\xi_{m+1})$.
Next, when the $(m+1)$-th subject is ready for randomization and the
corresponding covariate $\bkg\xi_{m+1}$ is recorded, we assign the
patient to treatment $1$ with a probability of
\begin{equation}\label{eqallocationDBCD}
\psi_{m+1,1}= \frac{\widehat{\pi}_m
\left(\frac{\widehat{\rho}_m}{N_{m,1}/m}\right)^{\gamma}}{
\widehat{\pi}_m
\left(\frac{\widehat{\rho}_m}{N_{m,1}/m}\right)^{\gamma}+
\left(1-\widehat{\pi}_m\right)
\left(\frac{1-\widehat{\rho}_m}{1-N_{m,1}/m}\right)^{\gamma}}
\end{equation}
 and to treatment $2$ with a probability of
$\psi_{m+1,2}=1-\psi_{m+1,1}$, where $\gamma\ge 0$ is a constant
that controls the degree of randomness of the procedure, from most
random when $\gamma=0$ to deterministic when $\gamma=\infty$. ZHCC's
design is a special case of CADBCD with $\gamma=0$.

 \bigskip
 \noindent{\bf Asymptotic properties.}~~
  For studying the asymptotic properties, we assume
the target allocation function $ \pi_1(\bm \theta^{\ast},\bm x)$
satisfies the following condition.

\begin{condition}\label{conditionA}
We assume that  the parameter space $\bm \Theta_k$ is a bounded
domain in $\mathbb R^d$, and that the true value $\bm \theta_k$ is an
interior point of $\bm \Theta_k$, $k=1,2$.
\begin{enumerate}
\item
For each fixed $\bm x$, $0<\pi_1(\bm \theta^{\ast},\bm x)<1$ is a
continuous function of $\bm \theta^{\ast}$ in the closure of $\bm
\Theta_1\times\bm \Theta_2$.
\item
 $\pi_1(\bm \theta^{\ast},\bm \xi)$
is twice differentiable with respect to $\bm \theta^{\ast}$, and the
expectations of $\|\partial \pi_1(\bm\theta,\bm\xi)/\partial \bm
\theta\|^2$ and
 $\sup\limits_{\|\bm\theta^{\ast}-\bm\theta\|\le \delta}\|\partial^2
 \pi_1(\bm\theta^{\ast},\bm\xi)/\partial \bm \theta^2\|$
 are finite for some $\delta>0$.
\end{enumerate}
\end{condition}
Write
$v=\ep [\pi_1(\bm \theta,\bm \xi)]$,  then $0<v<1$ due to Condition A.1.

\begin{theorem}\label{normalityCaDBCD} Suppose that for $k=1,2$,
\begin{eqnarray}\label{repoftheta}
\quad \widehat{\bm \theta}_{nk}-\bm \theta_k=\frac{1}{n}\sum_{m=1}^n
X_{m,k} \bm h_k(Y_{m,k},\bm \xi_m)\big(1+o(1)\big)+o(n^{-1/2})
\quad a.s.,
\end{eqnarray}
where $\bm h_k$s are
functions with $\ep[\bm h_k(Y_k,\bm\xi)|\bm\xi]=\bm 0$.
 We also
assume that $\ep\|\bm h_k(Y_k,\bm \xi)\|^2<\infty$, $k=1,2$.
Then under Condition \ref{conditionA}, we have
\begin{align}\label{eqConsiAssProb} \pr\big(X_{n,1} =1 \big)\to v; \quad
\pr\big(X_{n,1}=1|\mathscr F_{n-1}, \bm \xi_n=\bm x \big)\to \pi_1 (\bm \theta,
\bm x) \; a.s.
\end{align}
and
\begin{eqnarray}\label{eqLIL}
 \frac{ N_{n,1}}{n}-
v=O\Big(\sqrt{\frac{\log\log n}{n}}\Big) \; a.s.;\quad
 \widehat{\bm \theta}_n-\bm \theta=O\Big(\sqrt{\frac{\log\log
 n}{n}}\Big)\;\; a.s.
 \end{eqnarray}
Further, let
$ \bm V_k=\ep\{\pi_k(\bm \theta,\bm \xi)(\bm h_k(Y_k,\bm \xi))^{T}
\bm h_k(Y_k,\bm \xi)\}$,  $ k=1,2,$
 $ \bm V=diag\big(\bm V_1,  \bm V_2\big)$,
$\sigma_1^2=\ep[\pi_1(\bm\theta,\bm\xi)(1-\pi_1(\bm\theta,\bm\xi))]$,
$\sigma_2^2=\Var\{\pi_1(\bm\theta,\bm\xi)\}$,
$\sigma_3^2=\ep\frac{\partial\pi_1(\bm\theta,\bm\xi)}{\partial\bm
\theta} \bm V \big(\ep \frac{\partial \pi_1(\bm\theta,\bm\xi)
}{\partial\bm \theta}\big)^{T}$, $\lambda=\gamma
\frac{\sigma_1^2}{v(1-v)}$ and
$\sigma^2=\frac{\sigma_1^2+\sigma_3^2}{1+2\lambda}+\sigma_2^2+\sigma_3^2$.
Then,
\begin{eqnarray}\label{eqCLT} \sqrt{n}(  N_{n,1} / n-  v) \overset{D}\to N( 0,\sigma^2)
\; \text{ and } \; \sqrt{n}(\widehat{\bm \theta}_n-\bm \theta)
\overset{D}\to N(\bm 0,\bm V).
\end{eqnarray}
\end{theorem}

The proof of this Theorem is a little complex and will be state in
the Appendix. According to (\ref{eqConsiAssProb}), CADBCD satisfies
(\ref{property}). The asymptotic variability $\sigma^2$ of the
design takes the values from the maximum $2\sigma_3^2+v(1-v)$ when
$\gamma=0$ to the minim $\sigma_2^2+\sigma_3^2$ when
$\gamma=\infty$.

\bigskip
The next result for the generalized linear model
is a corollary of Theorem \ref{normalityCaDBCD}. The proof is given  in the Appendix through
the verification of Condition (\ref{repoftheta}).
\begin{corollary}  \label{thglm}
 Suppose the distributions of the responses  follow the generalized  linear model (\ref{eqglm}) and satisfy
  the following regular condition
\begin{equation}\label{eqregular}
H(\delta)=:\ep_{\bm\theta}\Bigl[\sup_{\|\bm z\|\le
\delta}\Bigl\|\frac{\partial^2\log
f_k(Y_k|\bm\xi,\bm\theta_k)}{\partial \bm\theta_k^2}
\biggr|_{\bm \theta_k}^{\bm\theta_k+\bm
z}\Bigr\|\Bigl]\to 0 \text{ as } \delta\to 0,
\end{equation}
where $f(x)|_a^b = f(b)-f(a)$. Under Condition \ref{conditionA}, if
the matrices $\bm I_1$ and $\bm I_2$ defined as in
(\ref{eqinformation1}) are nonsingular and the MLE
$\widehat{\bkg\theta}_m$, which maximize the likelihood function
(\ref{eqlikelihood}), is unique, then we have
(\ref{eqConsiAssProb}), (\ref{eqLIL}), and (\ref{eqCLT}) with $\bm
V=\bm I^{-1}(\bm\theta)$ and $\bm I(\bm\theta)=diag(\bm I_1,\bm
I_2)$.
\end{corollary}

It is obvious that $B(\bm\theta)=\sigma_2^2+\sigma_3^2$ is the best
asymptotic variability of CARA designs with two treatments according
to Definition \ref{definition1}. For the CADBCD,
$$ \sigma^2=\frac{\sigma_1^2+\sigma_3^2}{1+2\gamma\frac{\sigma_1^2}{v(1-v)}}
+B(\bm\theta)>B(\bm\theta)\;
 \text{but}\; \sigma^2\searrow B(\bm\theta) \text{ as } \gamma\nearrow\infty.$$
This means that the CADBCD is not asymptotically efficient but it
can approach an asymptotically efficient CARA design  if $\gamma$ is
chosen large. ZHCC's design is a special case of the CADBCD  which
has the largest variability.

\section{Conclusion Remarks} We have proposed a family of
covariate-adjusted response-adaptive designs that are fully
randomized and asymptotically efficient. The CADBCD can be viewed as
a generalization of Hu and Zhang's doubly adaptive biased coin
design (Hu and Zhang, 2004a) for incorporating covariate
information. The asymptotic properties derived here provide the
theoretical foundation for inference based on the CADBCD.

In this paper, we have assumed that the responses in each treatment
group are available without delay. In practice, there is no
logistical difficulty in incorporating delayed responses into the
CADBCD, provided that some responses become available during the
course of the allocation in the experiment, and thus we can always
update the estimates whenever new data become available. For
clinical trials with uniform (or exponential) patient entry and
exponential response times (see Bai, Hu and Rosenberger (2002),
Hu and Zhang (2004) and Zhang, et al (2006) for examples), it is easy to verify the
theoretical results in Section 2 and 3.

\section{Appendix: Proofs}

\noindent {\bf Proof of Theorem \ref{theorem1}.} Notice
$\ep[X_{m+1,k}|\mathscr{F}_m]=\ep[\psi_{m+1,k}|\mathscr{F}_m]\to
\rho_k(\bm\theta)$ by (\ref{property}) and $\{\sum_{m=1}^n (X_{m,k}-
\ep[X_{m,k}|\mathscr{F}_{m-1}]), \mathscr{F}_n\}$ is a martingale.
(\ref{eqtheorem1.1}) follows immediately. For (\ref{eqtheorem1.2}),
let $\mathscr{G}_m=\sigma(\mathscr{F}_m,\bm\xi_{m+1})$. Then
$\{\sum_{m=1}^n (X_{m,k}-
\ep[X_{m,k}|\mathscr{G}_{m-1}])I\{\bm\xi_m=x\}, \mathscr{G}_m\}$ is
a martingale with
$$\sum_{m=1}^n \ep\big[\left\{(X_{m,k}-
\ep[X_{m,k}|\mathscr{G}_{m-1}])I\{\bm\xi_m=x\}\right\}^2|\mathscr{G}_{m-1}\big]
\le N_n(x). $$ It follows that
$$\frac{\sum_{m=1}^n (X_{m,k}-
\ep[X_{m,k}|\mathscr{G}_{m-1}])I\{\bm\xi_m=x\}}{N_n(x)}\to 0\;\;
a.s. \;\text{ on }\; \{N_n(x)\to \infty\}$$
by Theorem 3.3.10 of Stout (1974).
On the other hand,
$$\frac{\sum_{m=1}^n (\ep[X_{m,k}|\mathscr{G}_{m-1}]-\pi_k(\bm\theta,x))I\{\bm\xi_m=x\}}{N_n(x)}\to 0\;\; a.s. \;\text{ on }\; \{N_n(x)\to \infty\}$$
by (\ref{property}). So, (\ref{eqtheorem1.2}) is proved. For
(\ref{eqtheorem1.3}), notice
$$\frac{N_n( B(\bm x,r))}{n}\to \pr\{\bm\xi\in B(\bm x,r)\}>0 \;\; a.s.
 $$
 With a similar argument we have
\begin{align*} \lim_{n\to\infty}
  \frac{N_{n,k|B(\bm x,r)}}{N_n(B(\bm x,r))} = & \lim_{n\to\infty}
  \frac{\sum_{m=1}^n \pi_k(\bm\theta,\bm\xi_m)I\{\bm\xi_m\in B(\bm x,r)\}}{N_n(B(\bm
  x,r))}\\
  =& \frac{\ep[\pi_k(\bm\theta,\bm\xi)I\{\bm\xi\in B(\bm x,r)\}]}{\pr\{\bm \xi\in B(\bm
  x,r)\}} \; \; a.s.
  \end{align*}
Letting $r\searrow 0$ yields (\ref{eqtheorem1.3}).

\bigskip

\noindent {\bf Proof of Theorem \ref{normalityCaDBCD}.} The proof is
a little complex and long. We will complete via four steps.

{\it Step 1.} We show that (\ref{eqLIL}) and
\begin{equation}\label{LILforRhoandN} \widehat{\rho}_m=v+O(\sqrt{\log\log m/m}) \; a.s.
\end{equation}

\smallskip Write $\pi_1=\pi_1(\bm\theta,\bm\xi)$ for short. Let
$M_{n,1}=\sum_{m=1}^n
\left(X_{m,1}-\ep[X_{m,1}|\mathscr{F}_{m-1},\bm\xi_m]\right)$,
$M_{n,2}=\sum_{m=1}^n\left(\pi_1(\bm\theta,\bm\xi_m)-\ep\pi_1\right)$,
$\bm Q_{n,k}=\sum_{m=1}^n X_{m,k} \bm h_k(Y_{m,k},\bm \xi_m)$ for
$k=1,2$, $\bm Q_n=(\bm Q_{n,1},\bm Q_{n,2})$ and $M_{n,3}=\bm
Q_n\left(\ep\frac{\partial \pi_1}{\partial \bm\theta}\right)^T$.
Then $\bm Q_n$ and $M_{n,j}$, $j=1,2,3$, are martingales. According
to the law of the iterated logarithm (LIL) for martingales, we have
\begin{equation}\label{eqLILforMart}
 \bm Q_n=O(\sqrt{\log\log n/n})  \; \text{ and }\;  M_{n,j}=O(\sqrt{\log\log n/n})\; a.s. j=1,2,3.
 \end{equation}
Hence, by (\ref{repoftheta}) it is easily shown that
\begin{equation}\label{eqLILfortheta}
\widehat{\bm\theta}_m-\bm \theta=O(\sqrt{\log\log m/m}) a.s.
\end{equation}
It follows that
\begin{align}\label{repofpi1}
 \widehat{\pi}_m=&\pi_1(\widehat{\bm\theta}_m,\bm\xi_{m+1})=\pi_1(\bm\theta,\bm\xi_{m+1})+(\widehat{\bm\theta}_m-\bm\theta)
\left(\frac{\partial \pi_1(\bm\theta,\bm\xi_{m+1})}{\partial \bm\theta}\right)^T\nonumber\\
&+O(1)\|\widehat{\bm\theta}_m-\bm\theta\|^2 \sup_{\|\bm\theta^{\ast}-\bm\theta\|\le \delta}
 \left\|\frac{\partial^2 \pi_1(\bm\theta^{\ast},\bm\xi_{m+1})}{\partial \bm\theta^2}\right\|\\
=&\pi_1(\bm\theta,\bm\xi_{m+1})+(\widehat{\bm\theta}_m-\bm\theta)
\ep\frac{\partial \pi}{\partial
\bm\theta}+(\widehat{\bm\theta}_m-\bm\theta) \left[\frac{\partial
\pi_1(\bm\theta,\bm\xi_{m+1})}{\partial \bm\theta}-\ep\frac{\partial
\pi_1}{\partial \bm\theta}\right]^T
\nonumber\\
&+O(1)\frac{\log\log m}{m}\sup_{\|\bm\theta^{\ast}-\bm\theta\|\le \delta}
 \left\|\frac{\partial^2 \pi_1(\bm\theta^{\ast},\bm\xi_{m+1})}{\partial \bm\theta^2}\right\| \;\; a.s.\nonumber
\end{align}
It is easily shown that
$$ \sum_{m=1}^n(\widehat{\bm\theta}_m-\bm\theta)
\left[\frac{\partial \pi_1(\bm\theta,\bm\xi_{m+1})}{\partial
\bm\theta}-\ep\frac{\partial \pi_1}{\partial \bm\theta}\right]^T
=o((\log n)^2) \;\; a.s. $$ and
$$ \sum_{m=1}^n \frac{\log\log m}{m}\sup_{\|\bm\theta^{\ast}-\bm\theta\|\le \delta}
 \left\|\frac{\partial^2 \pi_1(\bm\theta^{\ast},\bm\xi_{m+1})}{\partial \bm\theta^2}\right\|=o((\log n)^2) \;\; a.s. $$
 It follows that
\begin{equation}\label{repofsumpi}
 \sum_{m=1}^n\widehat{\pi}_m=\sum_{m=1}^n \pi_1(\bm\theta,\bm\xi_{m+1})
+\sum_{m=1}^n(\widehat{\bm\theta}_m-\bm\theta)
\left(\ep\frac{\partial \pi_1}{\partial \bm\theta}\right)^T+o((\log
n)^2) \; a.s.
\end{equation}
Similarly,
\begin{eqnarray}
& & \widehat{\rho}_m= \frac{1}{m}\sum_{i=1}^m \pi_1(\widehat{\bm\theta}_m,\bm\xi_i) \nonumber\\
 &=&\frac{1}{m}\sum_{i=1}^m \pi_1(\bm\theta,\bm\xi_i)
 +(\widehat{\bm\theta}_m-\bm\theta)\left(\ep\frac{\partial \pi_1(\bm\theta,\bm\xi)}{\partial \bm\theta}\right)^T\nonumber\\
 &&
 \label{repofrho1}+(\widehat{\bm\theta}_m-\bm\theta)
 \frac{1}{m}\sum_{i=1}^m\left[\frac{\partial \pi_1(\bm\theta,\bm\xi_i)}{\partial \bm\theta}-
 \ep\frac{\partial \pi_1(\bm\theta,\bm\xi)}{\partial \bm\theta}\right]^T\nonumber\\
 && \label{repofrho2}+O(1)\|\widehat{\bm\theta}_m-\bm\theta\|^2\frac{1}{m}\sum_{i=1}^m\sup_{\|\bm\theta^{\ast}-\bm\theta\|\le \delta}
 \left\|\frac{\partial^2 \pi_1(\bm\theta^{\ast},\bm\xi_i)}{\partial \bm\theta^2}\right\|
 \\
 &=&\frac{1}{m}\sum_{i=1}^m \pi_1(\bm\theta,\bm\xi_i)
 +(\widehat{\bm\theta}_m-\bm\theta)\left(\ep\frac{\partial \pi_1(\bm\theta,\bm\xi)}{\partial \bm\theta}\right)^T+O(\frac{\log\log m}{m}).
 \end{eqnarray}
It follows that
$$\widehat{\rho}_m= v+\frac{1}{m}\sum_{i=1}^m[ \pi_1(\bm\theta,\bm\xi_i)-\ep \pi_1]+ O(\sqrt{\log\log m/m})
=v+ O(\sqrt{\log\log m/m})\; a.s. $$ and
$$ \sum_{m=1}^n \widehat{\pi}_m =n v+ O(\sqrt{n \log\log n}) \; a.s. $$
Now, write
\begin{equation}\label{eqgfunction} g(\pi,a,b)=
 \frac{\pi(b/a)^{\gamma}}{\pi(b/a)^{\gamma}+(1-\pi)((1-b)/(1-a))^{\gamma}}.
 \end{equation}
 Then $\psi_{m+1,1}=g\left(\widehat{\pi}_m,
 N_{m,1}/m,\widehat{\rho}_m\right)$.
  It is easily seen that $g(\pi,a,b)$ is a non-decreasing function of $b$, and so $g(\pi,a,b)\le g(\pi,a,a)=\pi$ if $a\ge b$.
Let $l_n=\max\{m\le n: N_{m,1}/m\le \widehat{\rho}_m\}$, then
$\psi_{m+1,1}\le \widehat{\pi}_m$ when $m\ge l_n+1$. Hence
\begin{align}\label{eqproofLILN}
N_{n,1}=& N_{l_n+1,1}+M_{n,1}-M_{l_n+1,1} +\sum_{m=l_n+1}^{n-1}\psi_{m+1,1}\nonumber \\
\le & 1+N_{l_n,1}+M_{n,1}-M_{l_n+1,1}+\sum_{m=l_n+1}^{n-1}\widehat{\pi}_m \nonumber\\
\le & 1+ l_n \widehat{\rho}_{l_n}+M_{n,1}-M_{l_n+1,1}
+\sum_{m=1}^{n-1}\widehat{\pi}_m -\sum_{m=1}^{l_n}\widehat{\pi}_m\\
\le & nv +  O(\sqrt{n \log\log n})\;\; a.s.\nonumber
\end{align}
Similarly,
$$ n-N_{n,1}\le n(1-v)+  O(\sqrt{n \log\log n})\;\; a.s. $$
(\ref{eqLIL}) and (\ref{LILforRhoandN}) are now proved.

\medskip
{\it Step 2.} We show (\ref{eqConsiAssProb}) and the asymptotic
normality of $\widehat{\bm\theta}_n$.

\smallskip
 By (\ref{eqLIL}) and (\ref{LILforRhoandN}),
$\widehat{\rho}_n/(N_{n,1}/n)\to 1$ a.s.. And hence
(\ref{eqConsiAssProb}) is proved.
 and further
$\psi_{m,1}-\pi_1(\widehat{\bm\theta}_{m-1},\bm\xi_m)\to 0$ a.s.
Then, it is easily check that $\bm Q_n$ is a martingale with
\begin{align*}
&\frac{1}{n}\sum_{m=1}^n\ep\left[(\Delta \bm Q_n)^T\bm \Delta Q_n\right] \\
= & \frac{1}{n}\sum_{m=1}^n diag\Big(\ep\left[\psi_{m,1} \bm
h_1(Y_{m,1},\bm\xi_m)^T\bm h_1(Y_{m,1},\bm\xi_m)\right],\\
&\qquad\qquad \quad\;
\ep\left[\psi_{m,2} \bm h_2(Y_{m,2},\bm\xi_m)^T\bm h_2(Y_{m,2},\bm\xi_m)\right]\Big)\\
&\to \bm V.
\end{align*}
So, applying the central  limit theorem for martingales yields
$$ n^{1/2}(\widehat{\bm\theta}_n-\bm\theta)\overset{\mathscr{D}}\to N(\bm 0,\bm V). $$
The proof of Step 2 is completed.

\medskip
{\it Step 3.} We show that
\begin{equation}\label{repofpsi}
\psi_{m+1,1}=\widehat{\pi}_m-\gamma\frac{\widehat{\pi}_m(1-\widehat{\pi}_m)}{v(1-v)}
\left(\frac{N_{m,1}}{m}-\widehat{\rho}_m\right) +O(\frac{\log\log
m}{m}) \;\; a.s.
\end{equation}

\smallskip
 Let $g(\pi,a,b)$ be defined as in (\ref{eqgfunction}).
  By some elementary argument, it can be showed  that
\begin{equation}\label{expansionofg}
\sup_{0\le \pi\le 1} \left|g(\pi,a,b)-\pi+\gamma\frac{\pi
(1-\pi)}{v(1-v)} \left(a-b\right)\right|=O((a-v)^2+(b-v)^2),
\end{equation}
 as $(a,b)\to (v,v)$. By
(\ref{eqLIL}) and (\ref{LILforRhoandN}), it follows that
$$ \sup_{0\le \pi\le 1} \left|g(\pi,N_{m,1}/m,\widehat{\rho}_m)-\pi+\gamma\frac{\pi (1-\pi)}{v(1-v)}
\left(\frac{N_{m,1}}{m}-\widehat{\rho}_m\right)\right|=O(\frac{\log\log
m}{m}) \;\; a.s. $$
(\ref{repofpsi}) is now proved.

\medskip
{\it Step 4.} At last, we show the asymptotic normality of $N_n$.

\smallskip
Notice $N_{m,1}/m-\widehat{\rho}_m=O(\sqrt{\log\log m/m})$ a.s..
With the same argument as deriving (\ref{repofsumpi}), we can show
that
\begin{align*}
&\sum_{m=1}^n\frac{\widehat{\pi}_m(1-\widehat{\pi}_m)}{v(1-v)}
\left(\frac{N_{m,1}}{m}-\widehat{\rho}_m\right)\\
=&\sum_{m=1}^n\frac{\ep[\pi_1(1-\pi_1)]}{v(1-v)}
\left(\frac{N_{m,1}}{m}-\widehat{\rho}_m\right)+o((\log n)^2)\; a.s.
\end{align*}
By (\ref{repofpsi}) it follows that
\begin{align*}\sum_{m=1}^n \psi_{m-1,1}= &
\sum_{m=1}^n \pi_1(\bm\theta,\bm\xi_m)+\sum_{m=0}^{n-1}
(\widehat{\bm\theta}_m-\bm\theta) \left(\ep\frac{\partial
\pi_1}{\partial \bm\theta}\right)^T \\
& \quad -\lambda\sum_{m=1}^{n-1}
\left(\frac{N_{m,1}}{m}-\widehat{\rho}_{m}\right)+o((\log n)^2)\;
a.s.
\end{align*}
 Then
\begin{align*}
&N_{n,1}-nv= M_{n,1} +\sum_{m=1}^n \psi_{m-1,1}-n v \\
=& M_{n,1}+M_{n,2}+\sum_{m=0}^{n-1}
(\widehat{\bm\theta}_m-\bm\theta) \left(\ep\frac{\partial
\pi_1}{\partial \bm\theta}\right)^T \\
& -\lambda\sum_{m=1}^{n-1}
\left(\frac{N_{m,1}}{m}-\widehat{\rho}_{m}\right)+o((\log n)^2)\\
=& M_{n,1}+M_{n,2}+\lambda\sum_{m=1}^{n-1}\frac{M_{m,2}}{m}
+(\lambda+1)\sum_{m=0}^{n-1} (\widehat{\bm\theta}_m-\bm\theta)
\left(\ep\frac{\partial \pi_1}{\partial \bm\theta} \right)^T\\
& -\lambda\sum_{m=1}^{n-1}
\left(\frac{N_{m,1}}{m}-v\right)+o((\log n)^2)\; a.s.\\
=& M_{n,1}+\left(M_{n,2}+\lambda\sum_{m=1}^{n-1}\frac{M_{m,2}}{m}\right)
+(1+o(1))\left((\lambda+1)\sum_{m=1}^{n-1} \frac{ M_{m,3}}{m}\right)
 \\
& -\lambda\sum_{m=1}^{n-1}
\left(\frac{N_{m,1}}{m}-v\right)+o(n^{1/2})\; a.s.
\end{align*}
On the other hand,
$$  \ep[\Delta M_{m,i}\Delta M_{m,j}|\mathscr{F}_{m-1}]=0,\;\; i\ne j, $$
$$   \ep[(\Delta M_{m,1})^2|\mathscr{F}_{m-1}]
=\ep[\psi_{m,1}(1-\psi_{m,1})|\mathscr{F}_{m-1}]\to
\sigma_1^2\;\; a.s., $$
$$  \ep[(\Delta M_{m,2})^2|\mathscr{F}_{m-1}]= \Var[\pi_1(\bm\theta,\bm\xi_m)]= \sigma_2^2 $$
and
\begin{align*}
 \ep[(\Delta M_{m,3})^2|\mathscr{F}_{m-1}]
= \ep\frac{\partial \pi}{\partial \bm\theta} \ep[(\Delta
\bm Q_m)^T\Delta\bm Q_m|\mathscr{F}_{m-1}] \left(\ep\frac{\partial \pi}{\partial
\bm\theta}\right)^T
\to  \sigma_3^2 \;\; a.s.
\end{align*}
 By applying the function
central limit theorem (c.f., Corollary 3.1 of Hall and Heyde, 1980),
we have
$$ n^{-1/2}\left(M_{[nt],1},M_{[nt],2}, M_{[nt],3} \right)
\overset{\mathscr{D}}\to \left(\sigma_1B_t^{(1)}, \sigma_2
B_t^{(2)}, \sigma_3 B_t^{(3)}\right), $$ where $B_t^{(i)}$,
$i=1,2,3$, are three independent standard Brownian motions. Then
with the same argument as in Hu and Zhang (2004a), one can show that
$$ n^{-1/2}(N_{[nt],1}-[nt]v)\overset{\mathscr{D}}\to G_t, $$
where
$$ G_t=\sigma_1 t^{-\lambda} \int_0^t x^{\lambda} d B_x^{(1)}
+ \sigma_2 B_t^{(2)}+(\lambda+1)\sigma_3 t^{-\lambda} \int_0^t x^{\lambda-1}  B_x^{(3)}dx $$
 is a solution of the equation
$$ G_t= \sigma_1 B_t^{(1)} +\sigma_2 \left( B_t^{(2)} +\lambda \int_0^t \frac{B_x^{(2)}}{x} dx\right)
+(\lambda+1)\sigma_3 \int_0^t  \frac{B_x^{(3)}}{x} dx-\lambda \int_0^t \frac{G_x}x dx $$
with $G_0=0$. It is easily checked  that
$$ \Var(G_t)=t\left[ \frac{\sigma_1^2}{1+2\lambda}+\sigma_2^2+\frac{2(\lambda+1)}{1+2\lambda}\sigma_3^2\right]
=t\left[ \frac{\sigma_1^2+\sigma_3^2}{1+2\lambda}+\sigma_2^2+\sigma_3^2\right]. $$
Hence
$$ n^{1/2}(N_{n,1}/n-v)\overset{D}\to N(0,\sigma^2). $$


\bigskip

\noindent{\bf Proof of Corollary \ref{thglm}.} It is sufficient to show the strong continency of the MLE $\widehat{\bm\theta}_m$:
\begin{equation}\label{eqconofMLE}
\widehat{\bm\theta}_n\to \bm \theta.
\end{equation}
In fact, if (\ref{eqconofMLE}) is proved, then by (\ref{repofpi1})
and (\ref{repofrho1}) we have $\widehat{\rho}_n\to v$ a.s. and
$\frac{1}{n}\sum_{m=1}^n\widehat{\pi}_m \to v$ a.s.. By
(\ref{eqproofLILN}) we will have $N_n/n\to v$ a.s.  It follows that
$\psi_{m,k}-\pi_k(\widehat{\bm\theta}_{m-1},\bm\xi_{m})\to 0$ a.s. by (\ref{expansionofg}).
The rest proof is similar to Corollary 3.1 of Zhang et al (2007).

For (\ref{eqconofMLE}), it suffices to show that, for any
$\delta>0$ small enough, with probability one for $m$ large enough we have
 \begin{equation}\label{eqB1} \log
L_k(\bm \theta_k^{\ast})<\log L_k(\bm \theta_k), \text{ if }
\|\bm \theta_k^{\ast}-\bm \theta_k\|=\delta.
\end{equation}
We consider the case $k=1$ only. The application of Taylor's theorem yields
\begin{align*}
& \frac{1}{m}\log L_1(\bm \theta_1^{\ast})-\frac{1}{m}\log
L_1(\bm \theta_1)\nonumber\\
=&(\bm \theta_1^{\ast}-\bm \theta_1)\frac{1}{m}\frac{\partial \log
L_1}{\partial \bm \theta_1}\Big|_{\bm \theta_1}
+(\bm \theta_1^{\ast}-\bm \theta_1)\frac{1}{m}\frac{\partial^2 \log
L_1}{\partial
\bm \theta_1^2}\Big|_{\bm \theta_1}(\bm \theta_1^{\ast}-\bm \theta_1)^T
\nonumber\\
&
+(\bm \theta_1^{\ast}-\bm \theta_1)\Big\{\frac{1}{m}\int_0^1\Big[\left.\frac{\partial^2
\log L_1}{\partial \bm \theta_1^2}
\right|_{\bm \theta_1}^{\bm \theta_1+t(\bm \theta_1^{\ast}-
\bm \theta_1)}\Big]dt\Big\}(\bm \theta_1^{\ast}-\bm \theta_1)^T.
\end{align*}
Write
$$f(a,b,\bm z,\bm\xi)=\frac{\pi_1(\bm z,\bm\xi)\left(\frac{b}{a}\right)^{\gamma}}
{\pi_1(\bm z,\bm\xi)\left(\frac{b}{a}\right)^{\gamma}+(1-\pi_1(\bm
z,\bm\xi))\left(\frac{1-b}{1-a}\right)^{\gamma}}.$$ It is obvious
that $f$ is a continuous function of $a$, $b$ and $\bm z$ for each
give $\bm \xi$. By applying the law of large numbers for
martingales,  one can show that
$$\frac{1}{m}\frac{\partial \log
L_1}{\partial \bm \theta_1}\Big|_{\bm \theta_1}\to \bm 0\;\; a.s. $$
and
$$\frac{\partial^2 \log
L_1}{\partial
\bm \theta_1^2}\Big|_{\bm \theta_1}= \sum_{j=2}^m\big(\ep\big[f(a,b,\bm
z,\bm \xi)\bm  I_1(\bm \theta_1|\bm \xi)\big]\big)\Big|_{a=\frac{N_{j-1}}{j-1}, b=\widehat{\rho}_{j-1},\bm
z=\widehat{\bm \theta}_{j-1}}+o(m)\;\; a.s.$$
For the details of the proof, one can refer to Zhang et al (2007).
Further, it is obvious that
$$ \limsup \widehat{\rho}_m \le \lim \frac{1}{m}\sum_{j=1}^m \sup_{\bm\theta}\pi_1(\bm \theta,\bm\xi_j)
=\ep[ \sup_{\bm\theta}\pi_1(\bm \theta,\bm\xi)]<1 \;\; a.s., $$
where the superior is taken over the parameter space.   And
similarly
$$ \limsup_{n\to \infty} \frac{1}{n}\sum_{m=1}^n \widehat{\pi}_m\le \ep[ \sup_{\bm\theta}\pi_1(\bm \theta,\bm\xi)]<1 \;\; a.s. $$
By (\ref{eqproofLILN}),
$$ \limsup N_{n,1}/n\le \ep[ \sup_{\bm\theta}\pi_1(\bm \theta,\bm\xi)]<1 \;\; a.s. $$
By  considering $1-\widehat{\rho}_m$ and $n-N_{n,1}$ instead of
$\widehat{\rho}_m$ and $N_{n,1}$ respectively,  we have
$$ \liminf \widehat{\rho}_m \ge \ep[ \inf_{\bm\theta}\pi_1(\bm \theta,\bm\xi)]>0 \;\text{ and }\;
\liminf N_{n,1}/n \ge \ep[ \inf_{\bm\theta}\pi_1(\bm
\theta,\bm\xi)]>0\; a.s. $$
So we may assume that $\widehat{\rho}_m, N_{n,1}/n\in [\delta_0,1-\delta_0]$ for some $0<\delta_0<1$.
On the other hand, it is obvious that
$\bm y \ep\big[f(a,b,\bm z,\bm \xi)\bm  I_1(\bm \theta_1|\bm
\xi)\big]\bm y^T$ is a continuous function of $a,b,\bm y,\bm z$, and
is positive for all $0<a,b<1$, $\bm y\ne \bm 0$ and all $\bm z$. It
follows that there is a constant $c_0>0$ for which
$$ \liminf_{j\to \infty}\min_{\bm y:\|\bm y\|=1}\big(\bm y\ep\big[f(a,b,\bm
z,\bm \xi)\bm  I_1(\bm \theta_1|\bm \xi)\big]\bm y^T\big)_{a=\frac{N_{j-1}}{j-1}, b=\widehat{\rho}_{j-1},\bm
z=\widehat{\bm \theta}_{j-1}}>c_0\; a.s. $$
So  with probability one for $m$ large enough it holds that
\begin{align*}
& \frac{1}{m}\log L_1(\bm \theta_1^{\ast})-\frac{1}{m}\log
L_1(\bm \theta_1)\nonumber\\
 \le &
-\|\bm \theta_1^{\ast}-\bm \theta_1\|^2
\Big\{\frac{1}{m}\sum_{j=2}^m\min_{\bm y:\|\bm y\|=1}\big(\bm y\ep\big[f(a,b,\bm
z,\bm \xi)\bm  I_1(\bm \theta_1|\bm \xi)\big]\bm y^T\big)_{a=\frac{N_{j-1}}{j-1}, b=\widehat{\rho}_{j-1},\bm
z=\widehat{\bm \theta}_{j-1}}\Big\} \\
&+\|\bm \theta_1^{\ast}-\bm \theta_1\|^2
H(\|\bm \theta_1^{\ast}-\bm \theta_1\|)+o(1)\nonumber\\
\le& -c_0\delta^2+\delta^2H(\delta)+o(1)<0\; \text{uniformly in }
\bm \theta_1^{\ast} \text{ with }
\|\bm \theta_1^{\ast}-\bm \theta_1\|=\delta
\end{align*}
 when $\delta$ is small enough. (\ref{eqB1}) is proved.

\end{document}